\documentclass[sigconf]{acmart}
\usepackage{soul}
\usepackage{multirow}
\usepackage{enumitem}
\usepackage{makecell}

\AtBeginDocument{%
  \providecommand\BibTeX{{%
    \normalfont B\kern-0.5em{\scshape i\kern-0.25em b}\kern-0.8em\TeX}}}

\acmConference[Conference acronym 'XX]{Make sure to enter the correct
  conference title from your rights confirmation email}{June 03--05,
  2018}{Woodstock, NY}

\acmBooktitle{Woodstock '18: ACM Symposium on Neural Gaze Detection,
 June 03--05, 2018, Woodstock, NY} 
\acmISBN{978-1-4503-XXXX-X/18/06}

\begin{document}

\title{ABXI: Invariant Interest Adaptation for Task-Guided Cross-Domain Sequential Recommendation}

\author{Qingtian Bian}
\affiliation{%
  \institution{College of Computing and Data Science, Nanyang Technological University}
  \country{Singapore}}
\email{bian0027@e.ntu.edu.sg}

\author{Marcus Vinícius de Carvalho}
\affiliation{%
  \institution{College of Computing and Data Science, Nanyang Technological University}
  \country{Singapore}}
\email{marcus.decarvalho@ntu.edu.sg}

\author{Tieying Li}
\affiliation{%
  \institution{School of Computer Science and Engineering, Northeastern University}
  \country{China}}
\email{tieying@stumail.neu.edu.cn}

\author{Jiaxing Xu}
\affiliation{%
  \institution{College of Computing and Data Science, Nanyang Technological University}
  \country{Singapore}}
\email{jiaxing003@e.ntu.edu.sg}

\author{Hui Fang}
\authornote{Corresponding author}
\affiliation{%
  \institution{Research Institute for Interdisciplinary Sciences and Key Laboratory of Interdisciplinary Research of Computation and Economics, Shanghai University of Finance and Economics}
  \city{Shanghai}
  \country{China}}
\email{fang.hui@mail.shufe.edu.cn}

\author{Yiping Ke}
\affiliation{%
  \institution{College of Computing and Data Science, Nanyang Technological University}
  \country{Singapore}}
\email{ypke@ntu.edu.sg}

\begin{abstract}
Cross-Domain Sequential Recommendation (CDSR) has recently gained attention for countering data sparsity by transferring knowledge across domains. A common approach merges domain-specific sequences into cross-domain sequences, serving as bridges to connect domains. One key challenge is to correctly extract the shared knowledge among these sequences and appropriately transfer it. Most existing works directly transfer unfiltered cross-domain knowledge rather than extracting domain-invariant components and adaptively integrating them into domain-specific modelings. Another challenge lies in aligning the domain-specific and cross-domain sequences. Existing methods align these sequences based on timestamps, but this approach can cause prediction mismatches when the current tokens and their targets belong to different domains. In such cases, the domain-specific knowledge carried by the current tokens may degrade performance. To address these challenges, we propose the A-B-Cross-to-Invariant Learning Recommender (\textbf{ABXI}). Specifically, leveraging LoRA's effectiveness for efficient adaptation, ABXI incorporates two types of LoRAs to facilitate knowledge adaptation. First, all sequences are processed through a shared encoder that employs a domain LoRA for each sequence, thereby preserving unique domain characteristics. Next, we introduce an invariant projector that extracts domain-invariant interests from cross-domain representations, utilizing an invariant LoRA to adapt these interests into modeling each specific domain. Besides, to avoid prediction mismatches, all domain-specific sequences are aligned to match the domains of the cross-domain ground truths. Experimental results on three datasets demonstrate that our approach outperforms other CDSR counterparts by a large margin. The codes are available in \url{https://github.com/DiMarzioBian/ABXI}.
\end{abstract}



\maketitle

\begin{figure*}[t]
\centering
\includegraphics[width=1.0\linewidth]{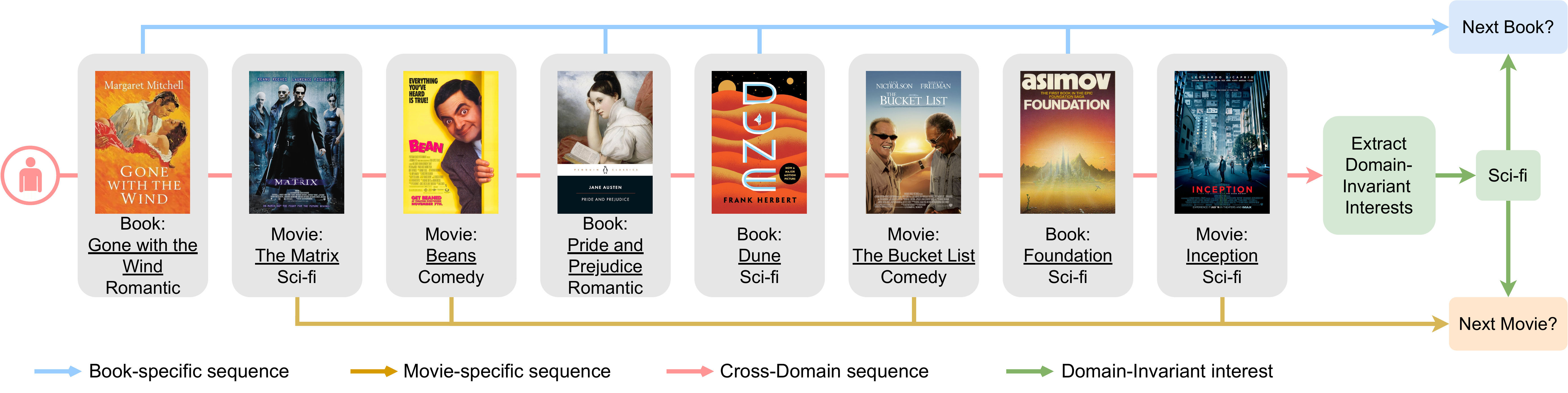}
\setlength{\abovecaptionskip}{-0.4cm}
\caption{Our proposal on generating recommendations by integrating domain-specific interests with domain-invariant interests extracted from the cross-domain sequence.}
\vspace{-0.4cm}
\label{fig: cdsr}
\end{figure*}
\vspace{-0.2cm}
\section{Introduction}
In the era of information explosion, the Internet is flooded with massive amounts of content, yet users are exposed to only a small fraction of it. Such data sparsity remains a persistent challenge in modern recommender systems. Cross-Domain Sequential Recommendation (CDSR) has recently emerged as a promising approach to alleviate this sparsity issue by transferring knowledge across different domains to enrich user profiles \cite{chen2024survey, alharbi2021cross, alharbi2022cross, lin2024mixed, cao2022contrastive, ye2023dream, xu2023multi, park2023cracking, zheng2022ddghm}.

A common strategy in CDSR involves merging domain-specific sequences into cross-domain sequences that serve as bridges, enabling mutual enhancement between domains \cite{alharbi2021cross, lin2024mixed, cao2022contrastive, ye2023dream, xu2023multi}. Figure~\ref{fig: cdsr} illustrates an example of a user's domain-specific and cross-domain interaction sequences. In the book domain, the user's interests encompass science fiction and romantic novels, while in the movie domain, the user prefers science fiction and comedy films. From the perspective of cross-domain sequences, the user's interest in science fiction can be leveraged in both the book and movie domains to create more comprehensive user profiles. On the contrary, the specific interests in romantic books and comedy movies should not be indiscriminately shared between domains. However, most existing CDSR approaches mix up the concepts of cross-domain and domain-invariant interests by directly transferring unfiltered cross-domain knowledge into domain-specific modeling. This practice can transfer information specific to a certain domain to interfere with others, adversely affecting recommendation performance. Therefore, extracting domain-invariant knowledge from cross-domain sequences is essential to facilitate effective sharing across specific domains.
\begin{figure*}[ht]
\centering
\includegraphics[width=0.9\linewidth]{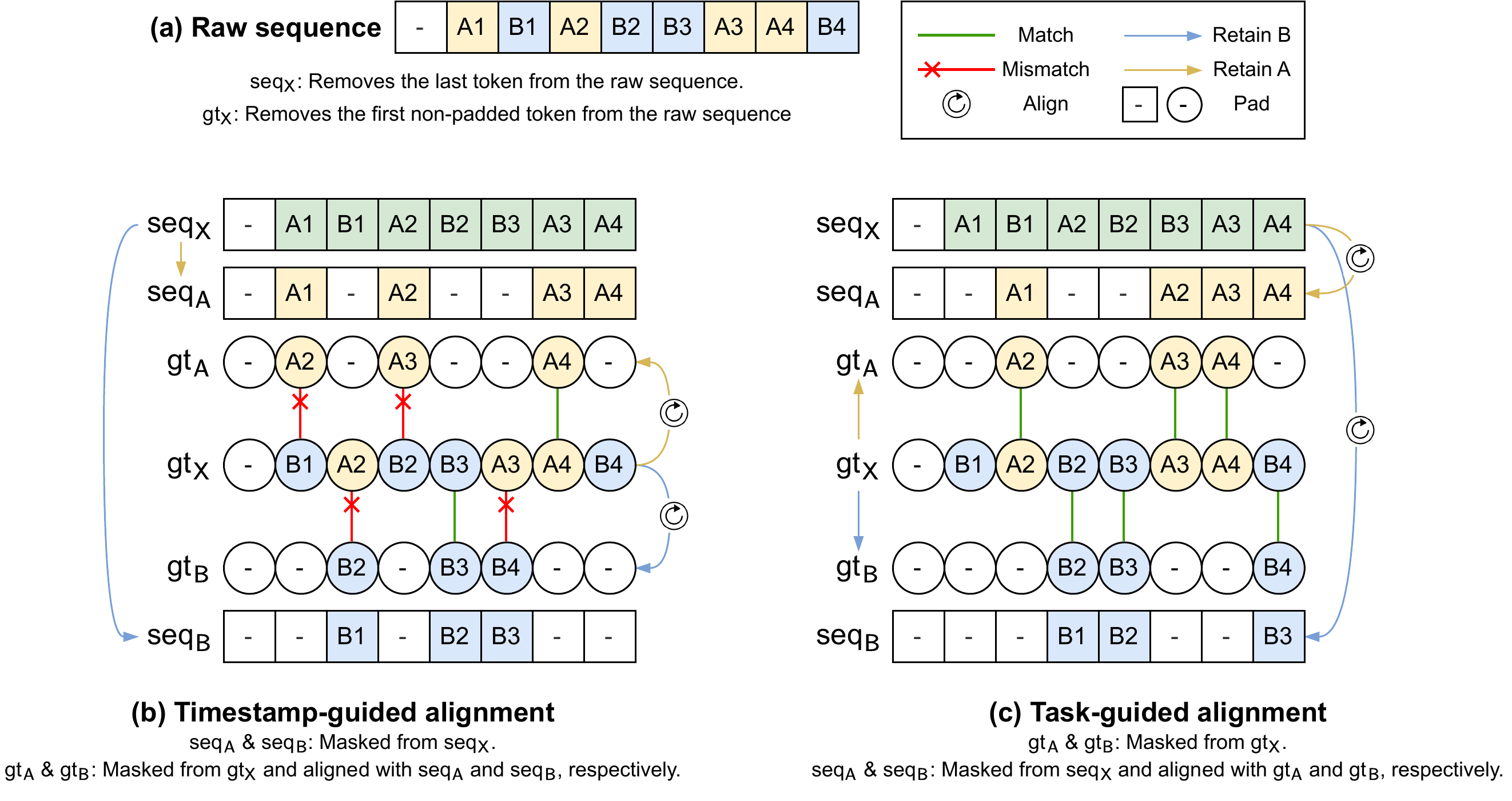}
\vspace{-0.3cm}
\setlength{\abovecaptionskip}{0.3cm}
\caption{Illustration of the sequence splits under different alignments, where \text{gt} denotes the ground truth. (a) illustrates the input raw sequence. (b) and (c) demonstrate the split outcomes of timestamp-guided and task-guided alignment, respectively.}
\vspace{-0.1cm}
\label{fig: seq}
\end{figure*}
Furthermore, another challenge lies in aligning domain-specific and cross-domain sequences when making recommendations within each domain. Current self-attention-based methods typically align cross-domain and domain-specific sequences based on timestamps \cite{cao2022contrastive, ye2023dream, lin2024mixed}, as depicted in Figure~\ref{fig: seq}b. Although this approach is intuitive and facilitates the fusion of cross-domain and domain-specific sequential features through direct token-wise addition, it has inherent limitations. Generally, cross-domain training sequences comprise a mixture of items from different domains. If one input token and its corresponding ground truth token belong to different domains, the domain-specific information encoded in this input token may negatively impact the prediction of the ground truth token. For instance, as illustrated in Figure~\ref{fig: seq}b, consider token B3 as the input token to predict token A4 in seq$_\mathsf{X}$. The timestamp-guided alignment enables the model to incorporate encoded cross-domain interests along seq$_\mathsf{X}$ as well as encoded domain-specific interests along seq$_\mathsf{A}$. However, since the target token B4 originates from domain $\mathsf{B}$, it does not correspond with the $\mathsf{A}$-specific knowledge, potentially degrading the model's performance. We refer to this issue as the prediction mismatch.

To tackle the prediction mismatch issue and address the challenges of exploiting domain-invariant interest, we propose the A-B-Cross-to-Invariant Learning Recommender (\textbf{ABXI}). Specifically, we first realign all domain-specific sequences according to the domains of the ground truths to prevent prediction mismatches with cross-domain sequences. We then employ a shared self-attention encoder as the sequence model to encode all sequences into sequential representations. This shared encoder deploys one domain LoRA (dLoRA) for each sequence, which can efficiently switch modes to encode every cross-domain and domain-specific sequence. Additionally, we instantiate an invariant projector to extract the domain-invariant interests from cross-domain representations. This projector has integrated one invariant LoRA (iLoRA) for recommendations in each specific domain to conduct efficient adaptation. While LoRAs are typically used for fine-tuning, we extend their application to single-stage training by concurrently training LoRA modules with all other components in ABXI. Having introduced these designs, ABXI renovates both the pipelines of obtaining cross-domain and domain-specific interests.

To thoroughly evaluate ABXI with state-of-the-art CDSR methods, we conduct extensive experiments on three publicly available datasets. Experimental results show that ABXI outperforms all baselines by a significant margin. Ablation studies and sensitivity analyses further demonstrate the effectiveness of our designs.

To conclude, our contributions can be summarized as follows:
\begin{itemize} 
    \item We identify the prediction mismatch problem within previous sequence-model-based CDSR works, and introduce a task-guided alignment to solve this problem.
    
    \item We introduce two types of LoRA: dLoRAs switch the mode of the encoder to handle encoding each sequence; iLoRAs adaptively integrate domain-invariant interests into recommendations in each specific domain. 
    
    \item Extensive experiments are provided to demonstrate the effectiveness of ABXI. Results show that ABXI outperforms all baselines including state-of-the-art CDSR counterparts.
\end{itemize}

The rest of this paper is organized as follows: Section~\ref{sec: related_work} provides an overview of related work. In Section~\ref{sec: method}, we formalize the CDSR problem we aim to solve and introduce our proposed ABXI. Section~\ref{sec: experiment} evaluates ABXI through extensive experiments; Section~\ref{sec: conclusion} presents the conclusion.

\begin{figure}[tp]
\centering
\includegraphics[width=\linewidth]{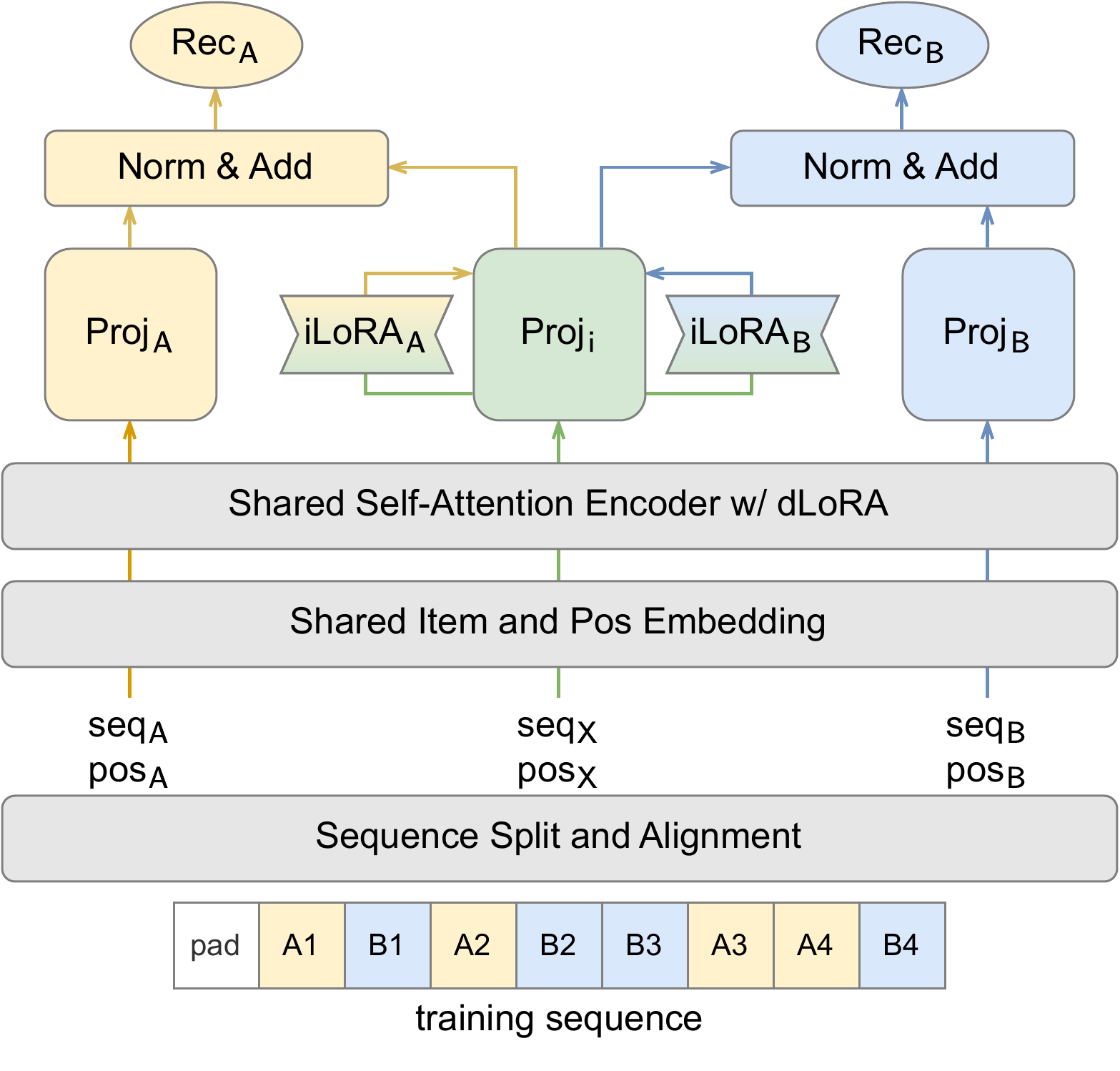}
\vspace{-0.4cm}
\setlength{\abovecaptionskip}{-0.1cm}
\caption{Proposed ABXI model.}
\vspace{-0.5cm}
\label{fig: model}
\end{figure}
%
\section{Related Work}\label{sec: related_work}
%
\subsection{Cross-Domain Recommendation}
Cross-Domain Recommendation (CDR) leverages transfer learning techniques to mitigate data sparsity. Common methods include domain alignment, which aligns users' or items' representations across different domains \cite{zhao2023cross, wang2021low, ma2024triple}, and domain adaptation, which adapts source knowledge to enhance target domains \cite{li2020ddtcdr, hu2018conet, cao2022disencdr, liu2020cross, zhu2022personalized}. Besides these typical CDR works, Multi-Modal Recommendation (MMR) can also be considered a form of CDR, as different modalities can be viewed as domains due to their shared semantics \cite{li2023hamur, wei2019mmgcn, wei2023multi, huang2019multimodal}.

Recently, Large Language Models (LLMs) have attracted much attention for their strong performance and scalability \cite{touvron2023llama, achiam2023gpt}. Researchers have attempted to introduce LLMs into recommender systems as well \cite{acharya2023llm, lin2024data}. However, these LLMs are Pretrained Language Models (PLMs) that are pretrained on Natural Language Processing (NLP) tasks. Therefore, researchers need to adapt these models to the recommendation domain. Under this perspective, using PLMs for recommendation can be seen as a cross-domain approach, where the source domain is the pretrained NLP domain, and the target domain is the recommendation domain.

A common way is to leverage Parameter-Efficient Fine-Tuning (PEFT) techniques to perform this adaptation at affordable costs. Most works utilize Low-Rank Adaptation (LoRA) \cite{hu2021lora} for such adaptation \cite{liao2024llara, acharya2023llm, lin2024data, zhao2024llm, yin2023heterogeneous}. Other techniques are also employed, such as prompt tuning \cite{penha2020does, yang2021improving, shen2023towards, zhang2023prompt} and adapter tuning \cite{fu2024exploring, hu2024enhancing}.

\subsection{Cross-Domain Sequential Recommendation}
Sequential Recommendation (SR) aims to predict users' next interacted items based on their historical interaction sequences \cite{kang2018self, sun2019bert4rec, bian2023cpmr}. Similar to CDR, CDSR introduces transfer learning into SR to conduct knowledge transfer in sequential scenarios. Early CDSR works focus on the assumption of multiple users sharing the same account \cite{ma2019pi, sun2021parallel, ma2022mixed, guo2021gcn}.

More recently, a broader concept of Cross-Domain Sequential Recommendation (CDSR) has emerged, focusing on leveraging bridging knowledge to enhance performance across domains. Depending on the type of knowledge transferred, CDSR approaches can be categorized into several types. Some studies \cite{lin2024mixed, li2022recguru, li2021dual, lin2024mixed} leverage overlapping users who have interacted in both domains as bridges to enhance performance for all users, including non-overlapping ones. Some methods \cite{xu2023multi, cao2022contrastive, ye2023dream, alharbi2021cross, alharbi2022cross, ding2023tpuf} focus exclusively on overlapping users to strengthen their profiling in target domains. IESRec \cite{liu2023joint} leverages semantic similarities in natural language to align domains in scenarios without overlapping users.

\section{Methodology}\label{sec: method}
%
\subsection{Problem Formulation}
In this study, we focus on the dual-target CDSR task, involving two distinct domains denoted as $\mathsf{A}$ and $\mathsf{B}$. Let a user's historical interaction sequences in domains $\mathsf{A}$ and $\mathsf{B}$ be represented as $\text{seq}_{\mathsf{A}} = (i_{\mathsf{A}1}, i_{\mathsf{A}2}, i_{\mathsf{A}3}, \dots, i_{\mathsf{A}n})$ and $\text{seq}_{\mathsf{B}} = (i_{\mathsf{B}1}, i_{\mathsf{B}2}, i_{\mathsf{B}3}, \dots, i_{\mathsf{B}m})$, respectively. The objective is to predict the user's next interaction items in each domain, specifically $i_{\mathsf{A}(n+1)}$ and $i_{\mathsf{B}(m+1)}$. The task can be formulated as:

\noindent\textbf{Input}: One user's domain-specific sequences, $\text{seq}_{\mathsf{A}} = (i_{\mathsf{A}1}, i_{\mathsf{A}2}, i_{\mathsf{A}3},\\ \dots, i_{\mathsf{A}n})$ and $\text{seq}_{\mathsf{B}} = (i_{\mathsf{B}1}, i_{\mathsf{B}2}, i_{\mathsf{B}3}, \dots, i_{\mathsf{B}m})$.

\noindent\textbf{Output}: A recommender system that estimates the probability of
this users' next items, $i_{\mathsf{A}(n+1)}$ and $i_{\mathsf{B}(m+1)}$, to interact.

\subsection{Overview}
The architecture of our proposed ABXI model, as depicted in Figure~\ref{fig: model}, comprises four components: (1) sequence formulation and embedding; (2) shared self-attention encoder with domain LoRA; (3) projectors with invariant LoRA; and (4) the optimization objective.

\subsection{Sequence Formulation and Embedding}
Given a user's raw training sequence, we first extract the last interaction and the first interaction to form the cross-domain sequence seq$_\mathsf{X}$ and the cross-domain ground truth gt$_\mathsf{X}$, following the seq2seq paradigm. For simplicity, we denote the combined domain as domain $\mathsf{X}$. As illustrated in Figure~\ref{fig: seq}c, we then derive the domain-specific ground truths gt$_\mathsf{A}$ and gt$_\mathsf{B}$ by masking the respective domains in gt$_\mathsf{X}$.

Subsequently, we create the domain-specific sequence seq$_\mathsf{A}$ and seq$_\mathsf{B}$ by masking the respective domain in seq$_\mathsf{X}$ and re-aligning them based on gt$_\mathsf{A}$ and gt$_\mathsf{B}$ with paddings. The aligned sequences seq$_\mathsf{A}$ and seq$_\mathsf{B}$ are ensured to have the same target items, position-wise, as seq$_\mathsf{X}$. In this way, the prediction mismatch issue encountered by previous works \cite{ye2023dream, cao2022contrastive} is addressed. 

Besides, we use gt$_\mathsf{A}$ and gt$_\mathsf{B}$ solely as intermediate terms to re-align domain-specific sequences, and they are not utilized during optimization. Consequently, the ground truths of seq$_\mathsf{X}$ and its seq$_\mathsf{A}$ and seq$_\mathsf{B}$ are unified, which eliminates the need for standalone domain-specific recommendation loss. The position indices of each sequence are assigned separately in reverse chronological order.

Each item is then embedded into a learnable vector at length $d$. We initialize the item embedding table $\mathbf{E}_{I}\in\mathbb{R}^{N_i\times d}$, where $N_i$ is the total number of items, and the position embedding table $\mathbf{E}_{P}\in\mathbb{R}^{L\times d}$, where $L$ is the maximum sequence length. Both embedding tables are shared across all sequences. Finally, The sequence embeddings for each sequence are obtained by adding the item and position embeddings, followed by a dropout operation to mitigate overfitting. We denote these sequence embeddings as $\boldsymbol{E}_{\mathsf{X}}$, $\boldsymbol{E}_{\mathsf{A}}$ and $\boldsymbol{E}_{\mathsf{B}}$ for domain $\mathsf{X}$, $\mathsf{A}$ and $\mathsf{B}$, respectively.

\subsection{Low-Rank Adaptation}
Many recommendation system studies \cite{guo2015trustsvd, ma2008sorec, jamali2010matrix, koren2008factorization} have already demonstrated that index-based methods exhibit low-rank natures because of data sparsity. LoRA \cite{hu2021lora}, designed as a PEFT technique, exploits similar low-rank characteristics in data to conduct efficient task adaptations \cite{{aghajanyan2020intrinsic}}. We leverage LoRA to conduct efficient domain adaptation in recommendation by proposing two modules: domain LoRA (\textbf{dLoRA}), which helps shared encoders adapt to both cross-domain and domain-specific modeling; and invariant LoRA (\textbf{iLoRA}), which adapts the extracted domain-invariant knowledge into specific final recommendations. Besides, notice that incorporating dropout operators can mitigate overfitting in LoRAs \cite{lin2024lora}; Therefore, the forward pass of our proposed LoRAs yields:
\begin{equation}\label{eq: lora}
\mathsf{LoRA}(\boldsymbol{X}) = {\boldsymbol{M}_B^{\uparrow}\boldsymbol{M}_A^{\downarrow}\boldsymbol{X}},
\end{equation}
where $\boldsymbol{M}_A^{\downarrow} \in \mathbb{R}^{r \times d}$, and $\boldsymbol{M}_B^{\uparrow} \in \mathbb{R}^{d \times r}$ denote the down- and up-projection matrices with rank $r < d$. We argue that LoRA's potential is not confined to fine-tuning but can be effectively applied in single-stage training as well. Thus, all LoRAs in ABXI are trained together with the rest of the model.

\subsection{Shared Encoder with Domain LoRAs}
Inspired by the effectiveness of SASRec \cite{kang2018self}, numerous sequential recommenders adopt self-attention encoder \cite{vaswani2017attention} as the backbone sequence model \cite{sun2019bert4rec, bao2023tallrec, petrov2023gsasrec}. Among them, all CDSR models instantiate multiple self-attention encoders for modeling sequences from different domains \cite{alharbi2022cross,alharbi2021cross, cao2022contrastive, ye2023dream, lin2024mixed}. 

We posit that a single self-attention encoder is sufficient to capture the majority of the necessary knowledge for recommendations, given the overlap in domain-invariant knowledge across domains. Consequently, we instantiate one shared encoder to all sequences. To preserve the specific uniqueness of each domain, we introduce three dLoRA modules in parallel with the encoder's feedforward network. These dLoRAs enable the shared encoder to switch modes efficiently among the domains $\mathsf{X}$, $\mathsf{A}$, and $\mathsf{B}$, thereby maintaining domain-specific nuances without compromising shared knowledge. The encoding process for a domain-$\mathsf{X}$ sequence embedding $\boldsymbol{E}_{\mathsf{X}}$ is formulated as follows:
\begin{equation}
\begin{gathered}
\boldsymbol{H}_{\mathsf{X}} = \mathsf{LN}\left(\boldsymbol{E}_{\mathsf{X}} + \mathsf{Drop}\left(\mathsf{MHA}\left(\boldsymbol{E}_{\mathsf{X}}\right)\right)\right), \\
\boldsymbol{H}_{\mathsf{X}}^{\mathsf{enc}} = \mathsf{LN}\left(\boldsymbol{H}_{\mathsf{X}} + \mathsf{Drop}\left(\mathsf{FFN}\left(\boldsymbol{H}_{\mathsf{X}}\right)\right) + \mathsf{Drop}\left(\mathsf{dLoRA_\mathsf{X}}\left(\boldsymbol{H}_{\mathsf{X}}\right)\right)\right),
\label{eq: encoder} 
\end{gathered}
\end{equation}
where $\mathsf{LN}$ denotes LayerNorm, $\mathsf{Drop}$ depicts dropout operator, $\mathsf{dLoRA_{X}}$ denotes the dLoRA unit for domain $\mathsf{X}$, $\mathsf{MHA}$ denotes the multi-head attention networks, and $\mathsf{FFN}$ denotes the feedforward networks. Similarly, by replacing the domain notation $\mathsf{X}$ with $\mathsf{A}$ and $\mathsf{B}$ in Eq.~\ref{eq: encoder}, we obtain the encoded sequential representations $\boldsymbol{H}_{\mathsf{A}}^{\mathsf{enc}}$ and $\boldsymbol{H}_{\mathsf{B}}^{\mathsf{enc}}$, respectively.

\subsection{Projectors with Invariant LoRAs}
To convert the encoded sequential representations into effective recommendation representations, we utilize a dedicated projector for each domain. Each domain-specific projector consists of a SwishGLU variant of MLP \cite{touvron2023llama}. The structure of this projector is defined as follows:
\begin{equation}
\mathsf{Proj}\left(\boldsymbol{X}\right) = \mathsf{Drop}\left( \left(\mathsf{Swish}\left(\boldsymbol{X}\boldsymbol{W}_1 \right)\otimes\boldsymbol{X}\boldsymbol{W}_2\right)\boldsymbol{W}_3 \right),
\end{equation}
where $\boldsymbol{W}_1\in\mathbb{R}^{d\times\frac{8}{3}d}$, $\boldsymbol{W}_2\in\mathbb{R}^{d\times\frac{8}{3}d}$ and $\boldsymbol{W}_3\in\mathbb{R}^{\frac{8}{3}d\times d}$ are learnable matrices. The Swish activation function is defined as $\mathsf{Swish}(x)=\frac{x}{1+e^{-\beta x}}$, with $\beta$ set to 1 \cite{touvron2023llama}.

For projectors within specific domains, we incorporate skip connections to obtain the projected domain-specific representations:
\begin{equation}
\begin{aligned}
\boldsymbol{H}_{\mathsf{A}}^{\mathsf{p}} = \mathsf{LN}\left(\boldsymbol{H}_{\mathsf{A}}^{\mathsf{enc}} + \text{Drop}\left(\text{Proj}_{\mathsf{A}}\left(\boldsymbol{H}^{\mathsf{enc}}_{\mathsf{A}}\right)\right)\right), \\
\boldsymbol{H}_{\mathsf{B}}^{\mathsf{p}} = \mathsf{LN}\left(\boldsymbol{H}_{\mathsf{B}}^{\mathsf{enc}} + \text{Drop}\left(\text{Proj}_{\mathsf{B}}\left(\boldsymbol{H}^{\mathsf{enc}}_{\mathsf{B}}\right)\right)\right).
\end{aligned}
\end{equation}

In contrast to domain-specific projectors, the domain-invariant projector Proj$_\mathsf{i}$ integrates two iLoRAs to adapt domain-invariant interests into final recommendations for domains $\mathsf{A}$ and $\mathsf{B}$. The projected invariant representations are obtained as follows:
\begin{equation}
\begin{aligned}
\boldsymbol{H}_{\mathsf{i2A}}^{\mathsf{p}} = \mathsf{LN}\left(\boldsymbol{H}_{\mathsf{X}}^{\mathsf{enc}} + \text{Drop}\left(\text{Proj}_{\mathsf{i}}\left(\boldsymbol{H}_{\mathsf{X}}^{\mathsf{enc}}\right)\right) + \text{Drop}\left(\mathsf{iLoRA_{A}}\left(\boldsymbol{H}_{\mathsf{X}}^{\mathsf{enc}}\right)\right)\right), \\
\boldsymbol{H}_{\mathsf{i2B}}^{\mathsf{p}} = \mathsf{LN}\left(\boldsymbol{H}_{\mathsf{X}}^{\mathsf{enc}} + \text{Drop}\left(\text{Proj}_{\mathsf{i}}\left(\boldsymbol{H}_{\mathsf{X}}^{\mathsf{enc}}\right)\right) + \text{Drop}\left(\mathsf{iLoRA_{B}}\left(\boldsymbol{H}_{\mathsf{X}}^{\mathsf{enc}}\right)\right)\right).
\end{aligned}
\end{equation}
Here, $\mathsf{Proj_i}$ shares the same structure as Proj$_\mathsf{A}$ and Proj$_\mathsf{B}$, while $\mathsf{iLoRA_{A}}$ and $\mathsf{iLoRA_{B}}$ are two instantiations of iLoRA.

The final recommendation representations are obtained by summing the projected invariant and domain-specific representations:
\begin{equation}
\begin{aligned}
\boldsymbol{H}_{\mathsf{A}}^\mathsf{rec} =\boldsymbol{H}_{\mathsf{A}}^\mathsf{p} + \boldsymbol{H}_{\mathsf{i2A}}^\mathsf{p}, \\
\boldsymbol{H}_{\mathsf{B}}^\mathsf{rec} = \boldsymbol{H}_{\mathsf{B}}^\mathsf{p} + \boldsymbol{H}_{\mathsf{i2B}}^\mathsf{p}.
\end{aligned}
\end{equation}

Moreover, the shared encoder dominates ABXI’s computational complexity, with projectors mirroring the FFN’s structure. The overall complexity is $\mathcal{O}(dL^2+Ld^2+dj)$, $j$ is the internal dimension of FFN. To extend ABXI to $N$ domains, simply instantiate one shared encoder, $N$ dLoRAs, $N$ iLoRAs, one domain-invariant projector, and $N$ domain-specific projectors.

\subsection{Optimization Objective}
Previous works' adopting timestamp-guided alignment \cite{ye2023dream, cao2022contrastive} requires separate optimization on the cross-domain sequences. However, since our proposed task-guided alignment unites the ground truths of domain-specific and cross-domain sequences, ABXI can be optimized entirely with only one set of positive and corresponding negative samples. We split the input training sequences in the manner of the seq2seq paradigm, and randomly select $N_\mathsf{neg}$ unobserved items within the same domain for each positive sample to form the negative set. We use InfoNCE \cite{DBLP:journals/corr/abs-1807-03748} to optimize ABXI. Given a recommendation representation $h$, we denote the embedding of the corresponding positive sample as $e^+$, and the embedding set of the union of positive and negative samples as $\boldsymbol{E}$. Therefore, the InfoNCE can thus be given as:
\begin{equation}
    f\left(h\right) = - \log\frac{\mathsf{exp}\left(h \cdot e^+/\tau\right)}{\sum_{e\in \boldsymbol{E}} \mathsf{exp}\left(h \cdot e/\tau\right)},
\end{equation}
where $\tau$ denotes the temperature factor. The final loss of sequence can then be obtained as:
\begin{equation}
    Loss = \frac{1}{|\boldsymbol{H}_{\mathsf{A}}^\mathsf{rec}|}\sum_{h\in\boldsymbol{H}_{\mathsf{A}}^\mathsf{rec}}f(h) + \frac{1}{|\boldsymbol{H}_{\mathsf{B}}^\mathsf{rec}|}\sum_{h\in\boldsymbol{H}_{\mathsf{B}}^\mathsf{rec}}f(h).
\end{equation}

%
\begin{table*}[!t]
\caption{Statistics of the CDSR datasets. GTs denote the number of ground truths.}
\vspace{-0.3cm}
\label{tab: data}
    \begin{tabular}{c|ccccccc}
    \toprule
    Dataset & \# Users & \# Items & \# Interactions & \# Val. GTs & \# Test GTs & \# $\mathsf{A}$$\rightarrow$$\mathsf{B}$ transitions & \# $\mathsf{B}$$\rightarrow$$\mathsf{A}$ transitions \\
    \midrule
    Food (A) & \multirow{2}{*}{7,144} & 11,837 & 83,663 & 2,837 & 2,419 & 30,308 & -\\
    Kitchen (B) & & 16,258 & 89,885 & 4,307 & 4,725 & - & 29,407 \\
    \midrule
    Beauty (A) & \multirow{2}{*}{4,474} & 10,379 & 50,329 & 2,086 & 1,875 & 17,841 & - \\
    Electronics (B) & & 14,188 & 63,800 & 2,388 & 2,599 & - & 17,888 \\
    \midrule
    Movie (A) & \multirow{2}{*}{28,350} & 35,712 & 347,65
     & 11,728 & 10,935 & 108,318 & - \\
    Book (B) & & 90,958 & 403,147 & 16,622 & 17,415 & - & 105,696 \\
    \bottomrule
    \end{tabular}
\vspace{-0.1cm}
\end{table*}
\begin{table}[t]
\caption{Hyperparameters selection.}
\vspace{-0.3cm}
\begin{tabular}{c|c}
    \toprule
    Hyperparameter & Value \\
    \midrule
    Embedding dimension $d$ & 256 \\
    Encoder layer & 1 \\
    Dropout rate & 0 to 0.7 in 0.1 increments \\
    Negative sample $N_{\text{neg}}$ & 128 \\
    Optimizer & AdamW \\
    Temperature $\tau$ & 0.75 \\
    Max / Warm-up epoch & 500 / 5\\
    Learning rate & \{$10^{-3}$, $10^{-4}$\} \\
    Weight decay & \{5, 2, 1\}$\times$\{10, 1, $10^{-1}$, $10^{-2}$, $10^{-3}$\}, 0 \\
    Learning rate decay & $\times 0.3162$, after 30 stable epochs \\
    Early-stopping patience & 60 epochs \\
    Random seed & \{3407, 0, 1, 2, 3\} \\
    \bottomrule
    \end{tabular}
\vspace{-0.3cm}
\label{tab: hyper}
\end{table}
%

\section{Experiments}\label{sec: experiment}
We design our experiments to answer the following research questions (RQ):
\begin{itemize}
    \item \textbf{RQ1}: How does ABXI perform in comparison to state-of-the-art CDSR models and other baseline methods?

    \item \textbf{RQ2}: How does the performance of ABXI change when the proposed task-guided alignment is replaced by previous timestamp alignment methods, or when the employed projectors, iLoRAs, or dLoRAs are removed?

    \item \textbf{RQ3}: What is the impact on ABXI's performance when the parameter-efficient iLoRA and dLoRA are replaced with dense layers?

    \item \textbf{RQ4}: How does the choice of the rank hyperparameter in iLoRA and dLoRA affect the performance of ABXI?
\end{itemize}

\subsection{Datasets}
We conduct our experiments on three datasets derived from the Amazon review datasets\footnote{\url{https://cseweb.ucsd.edu/~jmcauley/datasets/amazon/links.html}} \cite{mcauley2015image}, encompassing six distinct domains: Food-Kitchen (FK), Beauty-Electronics (BE), and Movie-Book (MB). Specifically, FK includes the `Grocery and Gourmet Food' as $\mathsf{A}$ and `Home and Kitchen' as $\mathsf{B}$; BE comprises `Beauty' as $\mathsf{A}$ and `Electronics' as $\mathsf{B}$; MB consists of `Movies and TV' as $\mathsf{A}$ and `Books' as $\mathsf{B}$. The HVIDEO dataset \cite{ma2019pi, sun2021parallel, ma2022mixed, guo2021gcn} is not utilized in this study because it focuses on shared-account CDSR. In this dataset, each account is accessed by multiple individuals, which fundamentally differs from the assumption in later CDSR works that a single user's interests across different domains can overlap and be shared. Similarly, the Douban dataset \cite{zhu2019dtcdr, zhu2020graphical} is excluded due to its coarse timestamp granularity limited to daily intervals. This limitation renders the order of interactions within the same day subjective, reducing reproducibility and potentially introducing biases.

In our preprocessing setup, each review is treated as a user interaction. We retain users who have interacted in both domains, aggregating and reordering their interactions chronologically based on the timestamps. Subsequently, we remove items that have been interacted with fewer than five times among these users. To further reduce computational load, we limit each user's interaction sequence to the latest 50 interactions, following \cite{kang2018self}. This truncation may result in some users no longer meeting the domain-overlapping criteria, necessitating a secondary filtering step to exclude these users.

The statistics of the processed datasets are summarized in Table~\ref{tab: data}. We evaluate all methods using five different random seeds to ensure the robustness and reproducibility of the results. Performance is measured using Hit Rate (HR), Normalized Discounted Cumulative Gain (NDCG), and Mean Reciprocal Rank (MRR). For single-target models, hyperparameters are selected based on the MRR score within each domain. In contrast, hyperparameters are selected for dual-target models based on the aggregate MRR scores across both domains.

\begin{table*}[t]
\caption{Overall performance (RQ1). The best and the runner-up are highlighted in \textbf{bold} and \underline{underlined} respectively. We use paired t-tests to assess the statistical significance of pairwise differences between ABXI and the best-performing baseline, with $p<0.01$ for all metrics.}
\vspace{-0.3cm}
  \begin{tabular}{p{5mm}c|cccc|cccc}
    \toprule
    \multirow{2}{*}{Type} & \multirow{2}{*}{Methods} & \multicolumn{4}{c|}{Food} & \multicolumn{4}{c}{Kitchen} \\
    & & HR@5 & HR@10 & NDCG@10 & MRR & HR@5 & HR@10 & NDCG@10 & MRR \\
    \midrule
    \multirow{5}{*}{ST} & SASRec-1 & 0.1930\tiny{±0.0028} & 0.2611\tiny{±0.0036} & 0.1561\tiny{±0.0021} & 0.1332\tiny{±0.0019} & 0.1241\tiny{±0.0026} & 0.1851\tiny{±0.0018} & 0.1040\tiny{±0.0005} & 0.0900\tiny{±0.0007} \\
    & BERT4Rec-1 & 0.1819\tiny{±0.0035} & 0.2528\tiny{±0.0037} & 0.1462\tiny{±0.0027} & 0.1230\tiny{±0.0030} & 0.1114\tiny{±0.0040} & 0.1685\tiny{±0.0036} & 0.0926\tiny{±0.0029} & 0.0810\tiny{±0.0025} \\
    & CD-SASRec & 0.1797\tiny{±0.0079} & 0.2454\tiny{±0.0046} & 0.1421\tiny{±0.0060} & 0.1197\tiny{±0.0066} & 0.1119\tiny{±0.0067} & 0.1757\tiny{±0.0070} & 0.0946\tiny{±0.0045} & 0.0821\tiny{±0.0039} \\
    & CD-ASR & 0.1976\tiny{±0.0042} & 0.2727\tiny{±0.0052} & 0.1616\tiny{±0.0028} & 0.1368\tiny{±0.0026} & 0.1345\tiny{±0.0043} & 0.1995\tiny{±0.0044} & 0.1107\tiny{±0.0037} & 0.0941\tiny{±0.0034} \\
    & MGCL & 0.1932\tiny{±0.0041} & 0.2673\tiny{±0.0054} & 0.1523\tiny{±0.0021} & 0.1260\tiny{±0.0018} & 0.1467\tiny{±0.0009} & 0.2157\tiny{±0.0026} & 0.1203\tiny{±0.0019} & 0.1017\tiny{±0.0019} \\
    \midrule
    \multirow{5}{*}{DT} & SASRec-2 & \underline{0.2313}\tiny{±0.0034} & 0.2854\tiny{±0.0049} & \underline{0.1797}\tiny{±0.0039} & \underline{0.1535}\tiny{±0.0034} & \underline{0.1510}\tiny{±0.0037} & \underline{0.2168}\tiny{±0.0049} & \underline{0.1248}\tiny{±0.0027} & \underline{0.1062}\tiny{±0.0021} \\
    & BERT4Rec-2 & 0.2223\tiny{±0.0053} & \underline{0.2956}\tiny{±0.0030} & 0.1727\tiny{±0.0033} & 0.1427\tiny{±0.0041} & 0.1363\tiny{±0.0043} & 0.2055\tiny{±0.0052} & 0.1116\tiny{±0.0032} & 0.0948\tiny{±0.0025} \\
    & C$^2$DSR & 0.1984\tiny{±0.0072} & 0.2574\tiny{±0.0116} & 0.1546\tiny{±0.0050} & 0.1311\tiny{±0.0035} & 0.1263\tiny{±0.0051} & 0.1879\tiny{±0.0061} & 0.1051\tiny{±0.0033} & 0.0903\tiny{±0.0027} \\
    & DREAM & 0.2158\tiny{±0.0043} & 0.2771\tiny{±0.0039} & 0.1698\tiny{±0.0025} & 0.1441\tiny{±0.0021} & 0.1377\tiny{±0.0021} & 0.2045\tiny{±0.0033} & 0.1138\tiny{±0.0012} & 0.0956\tiny{±0.0006} \\
    & ABXI & \textbf{0.2498}\tiny{±0.0022} & \textbf{0.3175}\tiny{±0.0041} & \textbf{0.1973}\tiny{±0.0025} & \textbf{0.1679}\tiny{±0.0028} & \textbf{0.1737}\tiny{±0.0031} & \textbf{0.2410}\tiny{±0.0026} & \textbf{0.1415}\tiny{±0.0012} & \textbf{0.1206}\tiny{±0.0014} \\
    \midrule
    \midrule
    & & \multicolumn{4}{c|}{Beauty} & \multicolumn{4}{c}{Electronics} \\
    & & HR@5 & HR@10 & NDCG@10 & MRR & HR@5 & HR@10 & NDCG@10 & MRR \\
    \midrule
    \multirow{5}{*}{ST} & SASRec-1 & 0.1837\tiny{±0.0014} & 0.2597\tiny{±0.0038} & 0.1523\tiny{±0.0020} & 0.1295\tiny{±0.0016} & 0.1345\tiny{±0.0052} & 0.1894\tiny{±0.0038} & 0.1111\tiny{±0.0032} & 0.0982\tiny{±0.0033} \\
    & BERT4Rec-1 & 0.1687\tiny{±0.0043} & 0.2438\tiny{±0.0043} & 0.1404\tiny{±0.0034} & 0.1197\tiny{±0.0032} & 0.1277\tiny{±0.0067} & 0.1832\tiny{±0.0069} & 0.1053\tiny{±0.0054} & 0.0930\tiny{±0.0050} \\
    & CD-SASRec & 0.1605\tiny{±0.0115} & 0.2530\tiny{±0.0166} & 0.1380\tiny{±0.0107} & 0.1162\tiny{±0.0085} & 0.1290\tiny{±0.0060} & 0.1842\tiny{±0.0030} & 0.1069\tiny{±0.0027} & 0.0948\tiny{±0.0032} \\ 
    & CD-ASR & 0.1661\tiny{±0.0065} & 0.2550\tiny{±0.0044} & 0.1424\tiny{±0.0033} & 0.1211\tiny{±0.0033} & 0.1355\tiny{±0.0049} & 0.1938\tiny{±0.0039} & 0.1122\tiny{±0.0033} & 0.0987\tiny{±0.0034} \\
    & MGCL & 0.1364\tiny{±0.0047} & 0.2109\tiny{±0.0078} & 0.1162\tiny{±0.0044} & 0.1001\tiny{±0.0035} & \underline{0.1537}\tiny{±0.0018} & 0.2159\tiny{±0.0042} & 0.1273\tiny{±0.0019} & \underline{0.1118}\tiny{±0.0012} \\
    \midrule
    \multirow{5}{*}{DT} & SASRec-2 & \underline{0.2292}\tiny{±0.0102} & \underline{0.3262}\tiny{±0.0032} & \underline{0.1866}\tiny{±0.0032} & \underline{0.1530}\tiny{±0.0032} & 0.1481\tiny{±0.0036} & 0.2169\tiny{±0.0041} & 0.1236\tiny{±0.0040} & 0.1058\tiny{±0.0039} \\
    & BERT4Rec-2 & 0.1970\tiny{±0.0027} & 0.3077\tiny{±0.0089} & 0.1679\tiny{±0.0057} & 0.1363\tiny{±0.0050} & 0.1519\tiny{±0.0049} & \underline{0.2246}\tiny{±0.0028} & \underline{0.1275}\tiny{±0.0031} & 0.1101\tiny{±0.0031} \\
    & C$^2$DSR & 0.1835\tiny{±0.0066} & 0.2645\tiny{±0.0034} & 0.1519\tiny{±0.0038} & 0.1290\tiny{±0.0037} & 0.1288\tiny{±0.0072} & 0.1859\tiny{±0.0063} & 0.1081\tiny{±0.0047} & 0.0960\tiny{±0.0043} \\
    & DREAM & 0.2090\tiny{±0.0047} & 0.3043\tiny{±0.0069} & 0.1742\tiny{±0.0032} & 0.1447\tiny{±0.0032} & 0.1216\tiny{±0.0040} & 0.1817\tiny{±0.0041} & 0.1023\tiny{±0.0024} & 0.0895\tiny{±0.0019} \\
    & ABXI & \textbf{0.2807}\tiny{±0.0082} & \textbf{0.3835}\tiny{±0.0050} & \textbf{0.2245}\tiny{±0.0043} & \textbf{0.1846}\tiny{±0.0038} & \textbf{0.1659}\tiny{±0.0021} & \textbf{0.2389}\tiny{±0.0032} & \textbf{0.1385}\tiny{±0.0014} & \textbf{0.1200}\tiny{±0.0019} \\
    \midrule
    \midrule
    & & \multicolumn{4}{c|}{Movie} & \multicolumn{4}{c}{Book} \\
    & & HR@5 & HR@10 & NDCG@10 & MRR & HR@5 & HR@10 & NDCG@10 & MRR \\
    \midrule
    \multirow{5}{*}{ST} & SASRec-1 & 0.2258\tiny{±0.0031} & 0.2961\tiny{±0.0037} & 0.1647\tiny{±0.0025} & \underline{0.1874}\tiny{±0.0027} & 0.1357\tiny{±0.0029} & 0.1789\tiny{±0.0033} & 0.1007\tiny{±0.0022} & 0.1147\tiny{±0.0023}\\
    & BERT4Rec-1 & 0.2329\tiny{±0.0018} & 0.3105\tiny{±0.0012} & 0.1927\tiny{±0.0007} & 0.1696\tiny{±0.0007} & 0.1638\tiny{±0.0017} & 0.2152\tiny{±0.0013} & 0.1378\tiny{±0.0008} & 0.1243\tiny{±0.0006} \\
    & CD-SASRec & 0.2347\tiny{±0.0022} & 0.3117\tiny{±0.0026} & 0.1940\tiny{±0.0015} & 0.1709\tiny{±0.0017} & \underline{0.1710}\tiny{±0.0042} & \underline{0.2253}\tiny{±0.0043} & \underline{0.1434}\tiny{±0.0030} & \underline{0.1285}\tiny{±0.0026} \\
    & CD-ASR & 0.2352\tiny{±0.0045} & 0.3052\tiny{±0.0032} & 0.1956\tiny{±0.0027} & 0.1743\tiny{±0.0024} & 0.1622\tiny{±0.0010} & 0.2118\tiny{±0.0024} & 0.1372\tiny{±0.0010} & 0.1244\tiny{±0.0010} \\ 
    & MGCL & 0.2097\tiny{±0.0042} & 0.2851\tiny{±0.0040} & 0.1726\tiny{±0.0032} & 0.1509\tiny{±0.0033} & 0.1248\tiny{±0.0038} & 0.1668\tiny{±0.0049} & 0.1043\tiny{±0.0035} & 0.0946\tiny{±0.0033} \\
    \midrule
    \multirow{5}{*}{DT} & SASRec-2 & 0.2303\tiny{±0.0046} & 0.3067\tiny{±0.0043} & 0.1903\tiny{±0032} & 0.1673\tiny{±0.0030} & 0.1356\tiny{±0.0015} & 0.1830\tiny{±0.0014} & 0.1146\tiny{±0.0011} & 0.1034\tiny{±0.0010} \\
    & BERT4Rec-2 & 0.2317\tiny{±0.0008} & 0.3095\tiny{±0.0011} & 0.1925\tiny{±0.0015} & 0.1697\tiny{±0.0017} & 0.1547\tiny{±0.0014} & 0.2063\tiny{±0.0012} & 0.1302\tiny{±0.0009} & 0.1176\tiny{±0.0009} \\
    & C$^2$DSR & 0.2299\tiny{±0.0019} & 0.3003\tiny{±0.0026} & 0.1911\tiny{±0.0010} & 0.1700\tiny{±0.0005} & 0.1316\tiny{±0.0050} & 0.1767\tiny{±0.0050} & 0.1123\tiny{±0.0032} & 0.1025\tiny{±0.0028} \\
    & DREAM & \underline{0.2507}\tiny{±0.0068} & \underline{0.3255}\tiny{±0.0044} & \underline{0.2082}\tiny{±0.0044} & 0.1848\tiny{±0.0043} & 0.1469\tiny{±0.0037} & 0.1973\tiny{±0.0037} & 0.1237\tiny{±0.0033} & 0.1118\tiny{±0.0031} \\
    & ABXI & \textbf{0.2859}\tiny{±0.0016} & \textbf{0.3682}\tiny{±0.0030} & \textbf{0.2388}\tiny{±0.0014} & \textbf{0.2118}\tiny{±0.0011} & \textbf{0.1973}\tiny{±0.0021} & \textbf{0.2571}\tiny{±0.0019} & \textbf{0.1669}\tiny{±0.0013} & \textbf{0.1502}\tiny{±0.0014} \\
    \bottomrule
  \end{tabular}
\label{tab: rq1}
\vspace{-0.cm}
\end{table*}
\begin{table*}[t]
\caption{MRR results from ablation studies (RQ2) and LoRA versus dense layers (RQ4).}
\vspace{-0.3cm}
  \begin{tabular}{ccccc|cc|cc|cc}
    \toprule
    & iLoRA & Proj & dLoRA & Alignment & Food & Kitchen & Beauty & Electronics & Movie & Book\\
    \midrule
    ABXI & \checkmark & \checkmark & \checkmark & task & \textbf{0.1679}\tiny{±0.0028} & 0.1206\tiny{±0.0014} & \textbf{0.1846}\tiny{±0.0038} & \textbf{0.1200}\tiny{±0.0019} & \underline{0.2118}\tiny{±0.0011} & 0.1502\tiny{±0.0014} \\
    \midrule
    V$_\mathsf{ts}$ & \checkmark & \checkmark & \checkmark & timestamp & 0.1290\tiny{±0.0044} & 0.0870\tiny{±0.0022} & 0.1389\tiny{±0.0028} & 0.0923\tiny{±0.0024} & 0.1803\tiny{±0.0024} & 0.1265\tiny{±0.0024} \\
    V1 & \checkmark & \checkmark & - & task & \underline{0.1673}\tiny{±0.0026} & 0.1182\tiny{±0.0011} & 0.1792\tiny{±0.0041} & 0.1177\tiny{±0.0021} & 0.2104\tiny{±0.0024} & 0.1488\tiny{±0.0007} \\
    V2 & \checkmark & - & \checkmark & task & 0.1648\tiny{±0.0048} & 0.1164\tiny{±0.0013} & 0.1792\tiny{±0.0050} & 0.1171\tiny{±0.0026} & 0.2096\tiny{±0.0023} & 0.1497\tiny{±0.0019} \\
    V3 & - & \checkmark & \checkmark & task & \textbf{0.1679}\tiny{±0.0024} & 0.1193\tiny{±0.0014} & \underline{0.1823}\tiny{±0.0057} & \underline{0.1193}\tiny{±0.0018} & 0.2101\tiny{±0.0023} & 0.1488\tiny{±0.0013} \\
    V4 & - & - & - & task & 0.1619\tiny{±0.0018} & 0.1106\tiny{±0.0024} & 0.1653\tiny{±0.0044} & 0.1164\tiny{±0.0019} & 0.1980\tiny{±0.0013} & 0.1356\tiny{±0.0007} \\
    \midrule
    V$_\mathsf{e3}$ & \checkmark & \checkmark & 3$\times$encoders & task & 0.1659\tiny{±0.0032} & 0.1198\tiny{±0.0011} & 0.1729\tiny{±0.0050} & 0.1112\tiny{±0.0022} & 0.2114\tiny{±0.0012} & 0.1488\tiny{±0.0012} \\
    V$_\mathsf{dp3}$ & \checkmark & \checkmark & 3$\times$Proj & task & 0.1668\tiny{±0.0018} & \underline{0.1208}\tiny{±0.0010} & 0.1729\tiny{±0.0050} & 0.1112\tiny{±0.0022} & 0.2114\tiny{±0.0012} & 0.1488\tiny{±0.0012} \\
    V$_\mathsf{ip3}$ & 3$\times\mathsf{Proj}$ & \checkmark & \checkmark & task & 0.1656\tiny{±0.0027} & \textbf{0.1209}\tiny{±0.0026} & 0.1781\tiny{±0.0055} & 0.1146\tiny{±0.0031} & 0.2102\tiny{±0.0027} & \textbf{0.1509}\tiny{±0.0004} \\
    V$_\mathsf{ip2}$ & 2$\times\mathsf{Proj}$ & \checkmark & \checkmark & task & 0.1643\tiny{±0.0023} & 0.1201\tiny{±0.0012} & 0.1790\tiny{±0.0055} & 0.1167\tiny{±0.0023} & \textbf{0.2120}\tiny{±0.0012} & \underline{0.1505}\tiny{±0.0014} \\
    \bottomrule
  \end{tabular}
\label{tab: rq34}
\vspace{-0.cm}
\end{table*}
%

\subsection{Baselines}
We compare ABXI with several baseline models, categorized into four types:

\begin{itemize}[leftmargin=1.5cm]
  \item[\textbf{ST-SDSR}] Single-Target Single-Domain Sequential Recommenders: SASRec-1 and BERT4Rec-1. 
  
  \item[\textbf{DT-SDSR}] Dual-Target Single-Domain Sequential Recommender: SASRec-2 and BERT4Rec-2. 
  
  \item[\textbf{ST-CDSR}] Single-Target Cross-Domain Sequential Recommender: CD-SASRec, CD-ASR and MGCL.
  
  \item[\textbf{DT-CDSR}] Dual-Target Cross-Domain Sequential Recommender: C$^2$DSR and DREAM.
\end{itemize}

\noindent These baseline models are described as follows:
\begin{itemize}[leftmargin=0.4cm]
  \item[$\bullet$] \textbf{SASRec} \cite{kang2018self} utilizes the self-attention encoder to generate sequential representations. We implement two versions: ST-SASRec-1 and SASRec-2, corresponding to ST and DT settings, respectively. Specifically, SASRec-2 computes the losses independently for each domain and then sums them, thereby preventing the model from biasing toward either domain.
  
  \item[$\bullet$] \textbf{BERT4Rec} \cite{sun2019bert4rec} introduce Cloze objectives on top of SASRec. Similarly, we use two versions: BERT4Rec-1 and BERT4Rec-2.
  
  \item[$\bullet$] \textbf{CD-SASRec} \cite{alharbi2022cross} aggregates the encoded source-domain sequences into the target-domain encoding using two self-attention encoders.
  
  \item[$\bullet$] \textbf{CD-ASR} \cite{alharbi2021cross} fuses source and target domain sequences encoded by separate self-attention encoders.
  
  \item[$\bullet$] \textbf{C$^2$DSR} \cite{cao2022contrastive} instantiate difference set of graphical and self-attention encoder to encode cross-domain and domain-specific sequences, leveraging augmentation for contrastive learning.
  
  \item[$\bullet$] \textbf{MGCL} \cite{xu2023multi} integrates graphical and sequential information under different views and strengthens the profiling via user-to-user contrastive learning on views.
  
  \item[$\bullet$] \textbf{DREAM} \cite{ye2023dream}  employs separate self-attention encoders for cross-domain and domain-specific sequences, incorporating specific-to-cross knowledge transferring and a similar user-to-user contrastive learning on domains.
  
\end{itemize}

\subsection{Implementation Details}
We adopt the leave-one-out strategy commonly used in SR. Specifically, we remove the last two interactions from each user sequence to serve as the ground truths for validation and testing. Evaluation metrics are computed separately for each domain. For easy and fair in evaluation, we follow the common strategy used in \cite{kang2018self, tang2018personalized, sun2019bert4rec}, sampling another 999 items from the same domain as the ground truth to serve as negative candidates. Besides basic statistics, Table \ref{tab: data} summarizes the number of validation and testing ground truths per domain, as well as the counts of cross-domain item-to-item transitions within all sequences (i.e., $\mathsf{A}$→$\mathsf{B}$ and $\mathsf{B}$→$\mathsf{A}$). Hyperparameters not specified elsewhere are listed in Table \ref{tab: hyper}.

Our experimental setup treats CDSR as SDSR with side domain information. By comparing both types of models under identical conditions, we aim to assess the performance improvements fairly. If new CDSR models do not outperform classic SDSR models within this setting, their practical utility may be limited. We conducted all experiments under AMD Ryzen Threadripper PRO 5995WX CPU and NVIDIA GeForce RTX 4090 GPU, leveraging PyTorch 2.2.1.

\subsection{Overall Performance Comparison (RQ1)}
The overall results are reported in Table \ref{tab: rq1}. Across all datasets, ABXI significantly outperforms all baseline models on all evaluation metrics including state-of-the-art CDSR models, with statistical significance ($\mathsf{p}$<0.01). This improvement is evident in both domains, which proves that a single sequence model with aid modules is sufficient to capture both cross-domain and domain-specific knowledge.

In contrast, other DT-CDSR models do not demonstrate substantial improvements over their SDSR counterparts. Only in the Movie domain do we observe that DREAM outperforms the SDSR models. This can be attributed to the prediction mismatches in timestamp-guided alignment within these DT-CDSR models. Conversely, ST- and DT-SDSR models do not encounter such issues, as they do not require domain-specific sequences. 

Furthermore, ST-CDSR models exhibit inferior performance to DT-CDSR models due to two key factors. First, our experimental setup does not intentionally create a sparse target domain, which limits the ability of ST-CDSR models to exploit their dense-to-sparse transfer capabilities. Second, existing ST-CDSR approaches directly transfer knowledge from the source domain to the target domain without proper filtration. Since only domain-invariant knowledge is appropriate for transfer, failing to isolate it from source-specific knowledge can result in negative transfer.

%
\subsection{Ablation Studies (RQ2)}
To further investigate the contribution of each proposed component, we design five ablation variants: \textbf{V}$\mathbf{_\mathsf{ts}}$ rollbacks to the timestamp-guided alignment; \textbf{V}$\mathbf{_1}$ removes all dLoRAs; \textbf{V}$\mathbf{_2}$ removes all projectors; \textbf{V}$\mathbf{_3}$ removes all iLoRAs; \textbf{V}$\mathbf{_4}$ removes all dLoRAs, iLoRAs and projectors. The results are reported in the middle section of Table~\ref{tab: rq34}. Compared to all ablation variants, ABXI achieves the best performance. Notably, we have the following observations.

Projectors contribute the most among all trainable modules. Due to their substantial number of learnable parameters, projectors provide sufficient capacity to transform encoded sequential representations into recommendation representations.

The performance gaps between V$\mathbf{_1}$ and ABXI across the three datasets demonstrate that dLoRA effectively enhances the encoding process with minimal additional learnable parameters. Similarly, the gaps between V$\mathbf{_3}$ and ABXI indicate that iLoRA facilitates the adaptation of domain-invariant knowledge to diverse downstream tasks. Furthermore, a comparison between V$\mathbf{_2}$ and V$\mathbf{_4}$ shows that incorporating parameter-efficient iLoRAs and dLoRAs significantly improves performance. While these LoRAs may not explicitly enhance performance in certain cases, careful rank tuning ensures that ABXI maintains performance comparable to models without them.

Among all variants, V$\mathbf{_5}$, which reverts to timestamp-guided alignment, exhibits the most significant performance degradation. This decline is due to prediction mismatches. Moreover, the absence of specialized domain-specific ground truths ABXI further exacerbates the deterioration.

\vspace{-0.2cm}
%
\subsection{LoRAs Versus Dense Layers (RQ3)}
LoRA is renowned for its parameter-efficient adaptation. However, in the context of non-LLM CDSR, we must determine whether this efficiency translates into effectiveness. Therefore, we design the following four variants: \textbf{V}$_\mathbf{e3}$: replace the shared encoder and dLoRAs with three self-attention encoders for domains $X$, $A$ and $B$, respectively; \textbf{V}$_\mathbf{dp3}$: replace three dLoRAs with three projectors; \textbf{V}$_\mathbf{ip3}$: replace $\mathsf{iLoRA_A}$ and $\mathsf{iLoRA_B}$ with $\mathsf{X}$-to-$\mathsf{A}$ and $\mathsf{X}$-to-$\mathsf{B}$ projectors; \textbf{V}$_\mathbf{ip2}$: remove the $\mathsf{Proj_i}$ in V$_\mathsf{i2}$. The results are reported in the bottom section of Table~\ref{tab: rq34}. 

Among the variants, V$_\mathsf{i3}$ and V$_\mathsf{i2}$ achieve closed performance to ABXI, suggesting redundancy between the $\mathsf{X}$-to-$\mathsf{A}$ and $\mathsf{X}$-to-$\mathsf{B}$ projectors. This indicates that their functionality can be effectively replaced by the combination of $\mathsf{Proj_i}$ and the two iLoRA modules.

V$_\mathsf{dp3}$, which replaces the parameter-efficient dLoRAs with dense MLP layers, results in a performance decline rather than an improvement. This highlights the superiority of LoRA in balancing effectiveness and efficiency. Similarly, V$_\mathsf{e3}$ degrades performance by instantiating three separate encoders for each domain instead of using a shared encoder with dLoRAs. This can be attributed to task-guided alignment, where domain-specific ground truths are unified into cross-domain ground truths for downstream recommendation tasks. Consequently, the domain-specific encoders end up learning the same knowledge as the cross-domain encoder, rendering them redundant.

\subsection{Rank Analysis in LoRA (RQ4)}
\begin{figure}[t]
\centering
\includegraphics[width=\linewidth]{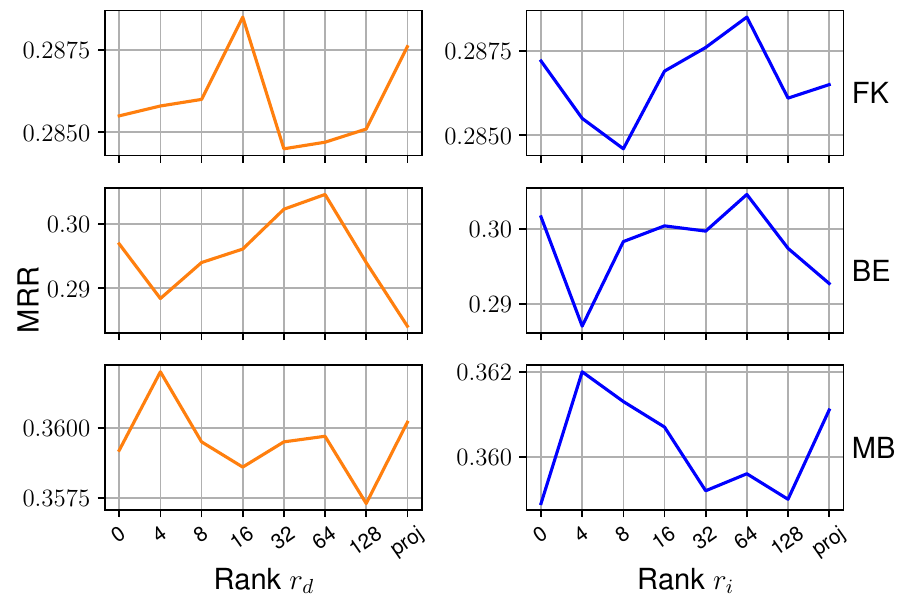}
\vspace{-0.3cm}
\setlength{\abovecaptionskip}{-0.cm}
\caption{MRR performance for varying rank values $r_d$ for dLoRA and $r_i$ for iLoRA (RQ4).}
\vspace{-0.4cm}
\label{fig: rank}
\end{figure}
This subsection examines the impact of different rank values $r_d$ for dLoRA and $r_i$ for iLoRA on ABXI's performance. We vary these ranks over \{0, 4, 8, 16, 32, 64, 128, proj\}, where $r=0$ denotes the removal of dLoRA (V1) or iLoRA (V3). The results are presented in Figure~\ref{fig: rank}. The tick label \textit{proj} in $r_d$ and $r_i$ plots corresponds to the variants V$_\mathsf{dp3}$ and V$_\mathsf{i3}$, which replace dLoRA and iLoRA, respectively, with projectors. This substitution is used to analyze the impact of the adaptation module's capacity on performance.

From Figure~\ref{fig: rank}, we observe that although the number of learnable parameters in LoRA is limited, it can largely impact performance across all domains, and inappropriate rank values may deteriorate performance. To achieve optimal results, we select $(r_d = 64, r_i = 64)$ for FK, $(r_d = 16, r_i = 64)$ for BE, and $(r_d = 4, r_i = 4)$ for MB dataset. While the original LoRA study \cite{hu2021lora} suggests that small ranks suffice for enhancing performance in NLP tasks, our experiments with ABXI reveal that larger ranks yield better results on FK and BE datasets. This difference can be attributed to the smaller size of these two datasets. In comparison, the MB dataset, which has significantly more data, performs better with smaller rankings of both LoRAs, which matches the preceding finding. Furthermore, the results for V$\mathsf{i3}$ show that indiscriminately increasing capacity hinders the learning of the domain-invariant projector Proj$\mathsf{i}$, resulting in performance degradation.

\section{Conclusion}\label{sec: conclusion} 
In this paper, we propose ABXI, a novel CDSR model. Specifically, ABXI addresses the prediction mismatch issue by integrating task-guided alignment to unify cross-domain and domain-specific optimizations. ABXI leverages domain projectors equipped with invariant LoRA modules to extract and adapt domain-invariant interests, enabling efficient and effective knowledge transfer Furthermore, ABXI employs a single shared encoder with domain LoRA to conduct efficient encoding. Extensive experimental results on three datasets demonstrated that ABXI significantly outperforms state-of-the-art CDSR models by a large margin. Ablation studies confirmed the effectiveness of each component, highlighting the importance of task-guided alignment and invariant interest extraction. For future work, we aim to further explore domain-invariant dynamics to achieve more accurate disentanglement.

\bibliographystyle{ACM-Reference-Format}
\balance
\bibliography{citation}


\end{document}